\documentclass[12pt]{article}
\def\real{\hbox{\rm I \hskip -6pt \bf R}}
\begin{document}
\title{Plane electromagnetic wave in PEMC}
\author{Bernard Jancewicz\\
Institute of Theoretical Physics, University of Wroc{\l}aw\\
pl. Maksa Borna 9, 50-204 Wroc{\l}aw, Poland,\\
Fax 48-71-3214454, Phone 48-71-3759412,\\
E-mail: bjan@ift.uni.wroc.pl}
\maketitle
\date{}
\begin{abstract}
Plane electromagnetic wave propagating in perfect electromagnetic
conductor (PEMC) is considered. Its wave number has no connection
with the frequency. An interface is introduced between an ordinary
isotropic medium and PEMC. The wave in PEMC is matched to plane
electromagnetic wave incident normally on the interface from the
ordinary medium and reflected from it. Then the plane-parallel
slab made of PEMC is considered and a plane wave is found in it.
\end{abstract}

\section{Introduction}

Recently Lindell and Sihvola \cite{ismo1} generalized the notions of perfect
electric conductor and perfect magnetic conductor to perfect electromagnetic
conductor (PEMC) for which the constitutive relation between the
electromagnetic fields is exotic in comparison with
the ordinary media like vacuum or air. In differential-form representation
of the fields this relation reads $G=\alpha F$ where $G=D-H\wedge dt$ and
$F=B+E\wedge dt$. The pseudoscalar parameter $\alpha $ was called admittance
in \cite{ismo2} and axion field in \cite{obuch}. Its zero limit yields the
perfect magnetic conductor, its infinity limit yields the perfect electric
conductor. For the three-dimensional fields {\bf E}, {\bf B}, {\bf H}, {\bf
D} the above constitutive relation means ${\bf D=\alpha B,~H=-\alpha E}$.

It is worth to consider the Maxwell energy-momentum tensor in
PEMC. The best appearance of this quantity for our purposes is the
energy-momentum 3-form [see \cite{hehl}, eq. (50)]:
\begin{equation} \Sigma _i=\frac{1}{2}(F\wedge e_i\rfloor
G-G\wedge e_i\rfloor F), \label{stress}\end{equation} where $e_i$
is the 4-dimensional vector basis of the tangent space. If one
substitutes $G=\alpha F$ with pseudoscalar $\alpha $ into this
expression, it yields zero. This means that for any
electromagnetic field in PEMC the energy density, energy flux and
the stress is zero. The same must be true for the electromagnetic
wave in such a medium.

In \cite{ismo1}, among others, a problem of linearly polarized plane
electromagnetic wave incident normally on a PEMC boundary was
considered by the method of duality transformation. It was found that the
reflected wave contains the cross-polarized term that is, the component of
the electric field perpendicular to that of the incident field. A similar
problem was considered in \cite{obuch}, namely the propagation of plane
electromagnetic wave in the ordinary medium with additional piecewise
constant axion field. It was shown that the reflection of the wave occurs at
an interface between two media with different axion values.

We present the explicit formula for a linearly polarized plane
electromagnetic wave in PEMC medium which -- as it was mentioned
above -- does not contain energy nor transmits energy. We consider
also the plane electromagnetic wave incident normally from the
vacuum on a boundary of PEMC and use the standard boundary
conditions to match it with the wave in PEMC. It turns out that
the reflected wave must be present on the side of the vacuum, and
it contains the cross-polarized term. It is an open question
whether the wave present in PEMC could be called transmitted wave,
if it does not transmit any energy.

Afterwards, we consider a plane-parallel slab made of PEMC and the
plane electromagnetic wave in it such that on the left-hand side
of the slab, the same incident and reflected waves are present,
and on the right-hand side, no wave is present.

\section{Plane electromagnetic wave}

In the differential-form formulation of electrodynamics the
electromagnetic fields are the following objects: {\bf B} is two-form, {\bf
D} is twisted two-form, {\bf E} is one-form, {\bf H} is twisted one-form.
The Maxwell equations are general, i.e. independent on the metric of space
and properties of a medium (we write them in a region devoid of charges and
currents):
\begin{equation} {\bf d\wedge B}=0,~~~~{\bf d\wedge E}+\frac{\partial {\bf
B}}{\partial t}=0. \label{maxw1}\end{equation}
\begin{equation} {\bf d\wedge D}=0,~~~~{\bf d\wedge H}-\frac{\partial {\bf
D}}{\partial t}=0, \label{maxw2}\end{equation}
The boundary conditions on a flat interface (a plane) $S$ without surface
charges and currents have the form
\begin{equation} ({\bf D}_1-{\bf D}_2)_{S}=0,~~~({\bf B}_1-{\bf B}_2)_{S}=0,
\label{bound1}\end{equation}
\begin{equation}({\bf E}_1-{\bf E}_2)_{S}=0,~~~({\bf H}_1-{\bf H}_2)_{S}=0,
\label{bound2}\end{equation} where the subscript $S$ denotes the
restriction of a given form to the plane $S$.\footnote{Restriction
of a differential form to $S$ means taking its values only on
vectors and bivectors parallel to $S$.} When the external form is
parallel to $S$, its restriction to $S$ is zero. For the
explanation of direction of an external form see \cite{janc1}. For
instance, in the Cartesian coordinates $x,y,z$ in flat space the
one-form ${\bf d}x$ is parallel to the $(Y,Z)$-plane and so on
with the cyclic change of variables, whereas the two-form ${\bf
d}x\wedge {\bf d}y$ is parallel to the $Z$-axis and so on.

We seek the solutions to the Maxwell equations in the form of
plane wave propagating along the $X$-axis, i.e. when the fields
are functions of a single scalar variable $\eta =\omega t-kx$,
called the {\it phase}:
\begin{equation} {\bf E}(x,t)=E_y(\eta )\,{\bf d}y+E_z(\eta )\,{\bf d}z,
\label{fala1}\end{equation}
\begin{equation} {\bf B}(x,t)=B_{xy}(\eta )\, {\bf d}x\wedge {\bf d}y+
B_{xz}(\eta )\,{\bf d}x\wedge {\bf d}z.
\label{fala2}\end{equation} Usually $E_y,\,E_z,\,B_{xy},\,B_{xz}$
are taken as combinations of sine and cosine functions of $\eta$
which is tantamount to assume that the wave is time-harmonic. We
present our reasoning without this assumption.

The Maxwell equation (\ref{maxw1})$_1$ is automatically satisfied,
because ${\bf d}B_{xy}$ and ${\bf d}B_{xz}$ are one-forms parallel
to ${\bf d}x$. We calculate ${\bf d\wedge E}=-kE'_y(\eta )\,{\bf
d}x\wedge {\bf d}y-kE'_z(\eta )\,{\bf d}x \wedge {\bf d}z$ and
substitute into (\ref{maxw1})$_2$:
$$-kE'_y(\eta )\,{\bf d}x\wedge {\bf d}y-kE'_z(\eta )\,{\bf d}x
\wedge {\bf d}z+\omega B'_{xy}(\eta )\, {\bf d}x\wedge {\bf d}y+\omega
B'_{xz}(\eta )\,{\bf d}x\wedge {\bf d}z=0,$$
$$(\omega B'_{xy}-kE'_y)\, {\bf d}x\wedge {\bf d}y+(\omega
B'_{xz}-kE'_z)\,{\bf d}x\wedge {\bf d}z=0.$$
Since the basic two-forms are linearly independent, we obtain
$$\omega B'_{xy}-kE'_y=0~~~~\hbox{and}~~~~\omega B'_{xz}-kE'_z=0.$$
Left-hand sides are functions of single variable $\eta$, hence
their integration yields
$$\omega B_{xy}-kE_y=\hbox{const},~~~~\omega
B_{xz}-kE_z=\hbox{const}.$$ It is natural to omit constant fields,
so the following proportionality is obtained
$$B_{xy}=\frac{k}{\omega}E_y~~~~\hbox{and}~~~~B_{xz}=\frac{k}{\omega}E_z,$$
for the functions of the scalar variable $\eta $. It follows from
this that
\begin{equation} {\bf B}=\frac{k}{\omega }\,{\bf d}x\wedge {\bf E}.
\label{fala3}\end{equation}
We similarly obtain for the two other fields
\begin{equation} {\bf H}(x,t)=H_x(\eta )\,{\bf d}y+H_y(\eta )\,{\bf d}z,
\label{fala4}\end{equation}
and
\begin{equation}{\bf D}=-\frac{k}{\omega }\,{\bf d}x\wedge {\bf H}.
\label{fala5} \end{equation}

\subsection{Plane wave in the conventional isotropic medium}

The constitutive relations for the isotropic medium with the electric
permittivity $\varepsilon $ and magnetic permeability $\mu $ have the form
\begin{equation} {\bf D}=\varepsilon *{\bf E},~~~{\bf B}=\mu *{\bf H},
\label{const1}\end{equation}
where $*$ denotes the Hodge star which maps the basic one-forms into basic
two-forms as follows:
\begin{equation} *{\bf d}x={\bf d}y\wedge {\bf d}z,~~~*{\bf d}y={\bf d}z\wedge
{\bf d}x,~~~*{\bf d}z={\bf d}x\wedge {\bf d}y.\label{hodge}\end{equation}
We treat here {\bf D} and {\bf H} as ordinary, non twisted forms, becasue
this does not influence the reasoning.

The relations (\ref{const1}) allow us to show that
$\frac{k}{\omega }= \sqrt{\varepsilon \mu}$ and to derive from
(\ref{fala1}) and (\ref{fala3}) two other electromagnetic fields
for the plane wave. We omit the standard reasoning in the
derivation. We now summarize the results and add the subscript $+$
to denote the fact that the planes of constant phase propagate in
the positive $X$-direction:
\begin{equation} {\bf E}_+(x,t)=E_{y+}(\eta _+)\,{\bf d}y+E_{z+}(\eta _+)
\,{\bf d}z,\label{fala1a}\end{equation}
\begin{equation} {\bf B}_+(x,t)=\sqrt{\varepsilon \mu }\,{\bf d}x\wedge
[E_{y+}(\eta _+)\,{\bf d}y+E_{z+}(\eta _+)\,{\bf d}z],
\label{fala2a}\end{equation}
\begin{equation} {\bf
H}_+(x,t)=\sqrt{\frac{\varepsilon}{\mu}}\,[-E_{z+}(\eta_+)\,{\bf
d}y+E_{y+}(\eta _+)\,{\bf
d}z],\label{fala3a}\end{equation}\begin{equation} {\bf
D}_+(x,t)=\varepsilon\,{\bf d}x\wedge [E_{z+}(\eta _+)\,{\bf
d}y-E_{y+}(\eta _+)\,{\bf d}z],\label{fala4a}\end{equation} where
$\eta _+=\omega t-kx$. We write also similar plane wave with the
phase propagating in the negative $X$-direction:
\begin{equation} {\bf E}_-(x,t)=E_{y-}(\eta _-)\,{\bf
d}y+E_{z-}(\eta _-)\,{\bf d}z,\label{fala1b}\end{equation}
\begin{equation} {\bf B}_-(x,t)=-\sqrt{\varepsilon \mu }\,{\bf d}x\wedge
[E_{y-}(\eta _-)\,{\bf d}y+E_{z-}(\eta _-)\,{\bf d}z],
\label{fala2b}\end{equation}
\begin{equation} {\bf
H}_-(x,t)=\sqrt{\frac{\varepsilon}{\mu}}\,[E_{z-}(\eta_-)\,{\bf
d}y-E_{y-}(\eta _-)\,{\bf d}z],\label{fala3b}\end{equation}\begin{equation}
{\bf D}_-(x,t)=\varepsilon\,{\bf d}x\wedge [E_{z-}(\eta _-)\,{\bf
d}y-E_{y-}(\eta _-)\,{\bf d}z],\label{fala4b}\end{equation}where $\eta
_-=\omega t+kx$.

\subsection{Plane wave in PEMC medium}

\medskip
We now write the constitutive relations for the PEMC medium
\begin{equation} {\bf H}=-\alpha {\bf E},
\label{pemc1}\end{equation}\begin{equation} {\bf D}=\alpha {\bf
B}. \label{pemc2}\end{equation}
They allow us to write immediately
the expressions for the fields ${\bf H}$ and ${\bf D}$ when
(\ref{fala1}) and (\ref{fala3}) are given.
%It is easy to see that (\ref{pemc1}) implies (\ref{pemc2}) when
%(\ref{fala3})and (\ref{fala5}) are satisfied.
Thus we summarize the formulas for the electromagnetic fields constituting
the plane electromagnetic wave propagating in the positive direction
of the $X$-axis:
\begin{equation} {\bf \tilde{E}}_+(x,t)=f_1(\xi _+)\,{\bf d}y+f_2(\xi
_+)\,{\bf d}z,\label{fala6}\end{equation}
\begin{equation} {\bf \tilde{B}}_+(x,t)=\frac{\tilde{k}}{\omega}\,{\bf d}x\wedge [f_1(\xi
_+)\,{\bf d}y+f_2(\xi_+)\,{\bf d}z], \label{fala7}\end{equation}
\begin{equation} {\bf \tilde{H}}_+(x,t)=-\alpha \,[f_1(\xi _+)\,{\bf d}y+f_2(\xi
_+)\,{\bf d}z], \label{fala8}\end{equation}
\begin{equation} {\bf \tilde{D}}_+(x,t)=\frac{\alpha
\tilde{k}}{\omega}\,{\bf d}x\wedge[f_1(\xi _+)\,{\bf d}y+f_2(\xi
_+)\,{\bf d}z], \label{fala9}\end{equation} where $\xi _+=\omega
t-\tilde{k}x$. There is no condition imposed on the quotient
$\frac{\tilde{k}}{\omega}$; this fact is different from the
situation known from the conventional medium.

The three fields ${\bf \tilde{B}}_+,\,{\bf \tilde{H}}_+,\,{\bf \tilde{D}}_+$
are parallel to ${\bf \tilde{E}}_+$ because the following relations are
satisfied:\begin{equation} {\bf \tilde{B}}_+=\frac{\tilde{k}}{\omega}{\bf
d}x\wedge {\bf \tilde{E}}_+,~~~{\bf \tilde{H}}_+=-\alpha {\bf
\tilde{E}}_+,~~~{\bf \tilde{D}}_+=\frac{\alpha \tilde{k}}{\omega}{\bf
d}x\wedge {\bf \tilde{E}}_+.\label{propor}\end{equation}
For this reason the energy density $w$ and the energy flux density {\bf S}
are zero
$$w=\frac{1}{2}\,({\bf \tilde{E}\wedge \tilde{D}+\tilde{H}\wedge
\tilde{B}})=0,~~{\bf S=\tilde{E}\wedge \tilde{H}}=0,$$ as expected
from the considerations in the Introduction.

Let us write down another plane wave propagating along the same axis with the
opposite phase velocity. The fields of this wave are expressed by
scalar functions $g_1$, $g_2$ of another scalar variable $\xi _-=\omega t+
\tilde{q}x$:
\begin{equation} {\bf \tilde{E}}_-(x,t)=g_1(\xi _-)\,{\bf d}y+g_2(\xi
_-)\,{\bf d}z,\label{fala10}\end{equation}
\begin{equation} {\bf \tilde{B}}_-(x,t)=-\frac{\tilde{q}}{\omega}\,{\bf
d}x\wedge [g_1(\xi _-)\,{\bf d}y+g_2(\xi_-)\,{\bf d}z],
\label{fala11}\end{equation}
\begin{equation} {\bf \tilde{H}}_-(x,t)=-\alpha
\,[g_1(\xi _-)\,{\bf d}y+g_2(\xi_-)\,{\bf d}z], \label{fala12}\end{equation}
\begin{equation} {\bf \tilde{D}}_-(x,t)=-\frac{\alpha
\tilde{q}}{\omega}\,{\bf d}x\wedge[g_1(\xi _-)\,{\bf d}y+g_2(\xi _-)\,{\bf
d}z], \label{fala13}\end{equation}
Again no condition is imposed on the quotient $\frac{\tilde{q}}{\omega} $.
We deliberately have chosen $\tilde{q}$ different from $\tilde{k}$. The
obvious assumption is that both are positive scalars.

\section{Reflection from PEMC boundary}

Let the space be divided on two parts by the plane $x=0$. For the left
half-space $x<0$ we assume homogeneous medium characterized by
$\alpha =0$ and constant values $\varepsilon , \mu$. For the right half-space
$x>0$ we assume $\varepsilon =\mu =0$ and constant value $\alpha $ which
means that it is PEMC medium. Consider
a plane linearly polarized electromagnetic wave travelling along the $X$-axis
in the left half-space. Such a normally incident wave will partially penetrate
the PEMC medium and partially be reflected from the interface. Therefore we
assume that in the left medium the electromagnetic field will be a
superposition of plane waves, right- and left-travelling along the $X$-axis.

We assume that the right-travelling wave has the linear
polarization parallel to $Y$-axis; such a field is expressed as
follows
\begin{equation} {\bf E}_+(x,t)=E_{y+}(\eta _+)\,{\bf
d}y,\label{fala18}\end{equation}
\begin{equation} {\bf B}_+(x,t)=\sqrt{\varepsilon \mu}\,{\bf
d}x\wedge E_{y+}(\eta _+)\,{\bf d}y,\label{fala19}\end{equation}
\begin{equation} {\bf H}_+(x,t)=\sqrt{\frac{\varepsilon }{\mu}}\,
E_{y+}(\eta _+)\,{\bf d}z,\label{fala20}\end{equation}
\begin{equation} {\bf D}_+(x,t)=-\varepsilon {\bf d}x\wedge E_{y+}(\eta _+)
\,{\bf d}z.\label{fala21}\end{equation}

One cannot expect that the reflected wave will have the same
linear polarization -- in fact, it has to contain a component of
{\bf E} parallel to ${\bf d}z$, because this gives rise to a
component of {\bf H} parallel to ${\bf d}y$ which by (\ref{pemc1})
must be present in the right medium. Thence we admit that the
reflected wave in the left medium has the general form
(\ref{fala1b}--\ref{fala4b}). Thus the full electromagnetic field
in the left medium is the following superposition
\begin{equation} {\bf E}(x,t)=[E_{y+}(\eta _+)+E_{y-}(\eta _-)]\,{\bf d}y+
E_{z-}(\eta _-) \,{\bf d}z,\label{fala22}\end{equation}
\begin{equation} {\bf B}(x,t)=\sqrt{\varepsilon \mu }\,{\bf d}x\wedge
\{ [E_{y+}(\eta _+)-E_{y-}(\eta _-)]\,{\bf d}y
-E_{z-}(\eta _-)\,{\bf d}z\}, \label{fala23}\end{equation}
\begin{equation} {\bf H}(x,t)=\sqrt{\frac{\varepsilon}{\mu}}\,\{ E_{z-}(\eta
_-)\,{\bf d}y+[E_{y+}(\eta _+)-E_{y-}(\eta _-)\,{\bf d}z]\},
\label{fala24}\end{equation}
\begin{equation} {\bf D}(x,t)=\varepsilon\,{\bf d}x\wedge \{E_{z-}(\eta _-)
\,{\bf d}y-[E_{y+}(\eta _+)+E_{y-}(\eta _-)\,{\bf d}z],
\label{fala25}\end{equation}

In order to not restrict generality we assume that in the right medium the
electromagnetic field is also a superposition of two opposite travelling
plane waves, hence we write the sums of expressions
(\ref{fala6}--\ref{fala9}) with the corresponding expressions
(\ref{fala10}--\ref{fala13})
\begin{equation} {\bf \tilde{E}}(x,t)=[f_1(\xi _+)+g_1(\xi _-)]\,{\bf d}y
+[f_2(\xi _+)+g_2(\xi _-)]\,{\bf d}z,
\label{fala26}\end{equation}
\begin{equation} {\bf \tilde{B}}(x,t)=\frac{1}{\omega}\,{\bf d}x\wedge
\{ [\tilde{k}f_1(\xi_+)-\tilde{q}g_1(\xi _-)]\,{\bf d}y+[\tilde{k}f_2(\xi_+)
-\tilde{q}g_2(\xi _-)]\,{\bf d}z]\},
\label{fala27}\end{equation}
\begin{equation} {\bf \tilde{H}}(x,t)=-\alpha \,\{ [f_1(\xi _+)+g_1(\xi _-)]
\,{\bf d}y+[f_2(\xi_+)+g_2(\xi _-)]\,{\bf d}z\}, \label{fala28}\end{equation}
\begin{equation} {\bf \tilde{D}}(x,t)=\frac{\alpha}{\omega}
\,{\bf d}x\wedge \{ [\tilde{k}f_1(\xi _+)-\tilde{q}g_1(\xi
_-)]\,{\bf d}y+ [\tilde{k}f_2(\xi _+)-\tilde{q}g_2(\xi _-)]\,{\bf
d}z\} , \label{fala29}\end{equation} We now consider the boundary
conditions (\ref{bound1}), (\ref{bound2}) on the plane $x=0$. The
conditions (\ref{bound1}) are satisfied trivially, because the
two-forms (\ref{fala23}), (\ref{fala25}), (\ref{fala27}) and
(\ref{fala29}) are parallel to the interface (they contain the
factor ${\bf d}x$), thus their restrictions to it are zero. The
one-forms (\ref{fala22},\ref{fala24},\ref{fala26},\ref{fala28})
are perpendicular to the interface, hence their restrictions to it
are equal to themselves. Thus the boundary conditions
(\ref{bound2}) reduce to
$${\bf E}(0,t)={\bf \tilde{E}}(0,t),~~~~~{\bf H}(0,t)={\bf \tilde{H}}(0,t),$$
and, after equating independent components, yield four equalities
\begin{equation} E_{y+}(\omega t)+E_{y-}(\omega t)=f_1(\omega t)+g_1(\omega
t), \label{war1}\end{equation}
\begin{equation} E_{z-}(\omega t)=f_2(\omega t)+g_2(\omega t),
\label{war2}\end{equation}
\begin{equation} \sqrt{\frac{\varepsilon}{\mu}}\,E_{z-}(\omega t)=
-\alpha [f_1(\omega t)+g_1(\omega t)], \label{war3}\end{equation}
\begin{equation} \sqrt{\frac{\varepsilon}{\mu}}\,[E_{y+}(\omega
t)-E_{y-}(\omega t)]=-\alpha \,[f_2(\omega t)+g_2(\omega t)].
\label{war4}\end{equation} It is interesting to notice that $f_i$
and $g_i$ appear only in sums $f_i(\omega t)+g_i(\omega t)$, hence
the field present in PEMC medium can be chosen in arbitrary
combination of the right- and left-travelling waves. By
eliminating $f_i+g_i$ from the above equations we arrive at the
following two linear equations
$$\sqrt{\frac{\varepsilon}{\mu}}\,E_{z-}=-\alpha (E_{y+}+E_{y-}),$$
$$\sqrt{\frac{\varepsilon}{\mu}}\,(E_{y+}-E_{y-})=-\alpha E_{z-},$$
which allow us to express the components $E_{y-}$ and $E_{z-}$ of the
reflected wave by the component $E_{y+}$ of the incident wave:
\begin{equation} E_{y-}=\frac{\varepsilon -\alpha ^2\mu}{\varepsilon +\alpha
^2\mu}\,E_{y+}, ~~~~~~~~ E_{z-}=-\frac{2\alpha \sqrt{\varepsilon
\mu}}{\varepsilon +\alpha ^2\mu}\,E_{y+}.
\label{war6}\end{equation} We insert this into (\ref{fala1b}) and
(\ref{fala3b}) to obtain the field strengths of the reflected
wave:
\begin{equation} {\bf E}_-(x,t)=\frac{1}{\varepsilon +\alpha
^2\mu}\,E_{y+}(\eta _-)\,[(\varepsilon -\alpha ^2\mu)\,{\bf d}y-2\alpha
\sqrt{\varepsilon \mu}\,{\bf d}z], \label{refl1}\end{equation}
\begin{equation} {\bf H}_-(x,t)=-\frac{\sqrt{\varepsilon /\mu}}{\varepsilon
+\alpha ^2\mu}\,E_{y+}(\eta _-)\,[2\alpha \sqrt{\varepsilon \mu}
\,{\bf d}y +(\varepsilon -\alpha ^2\mu)\,{\bf d}z].
\label{refl2}\end{equation} Formula (\ref{refl1}) coincides (after
appropriate change of notation) with the formula (41) in
\cite{ismo1}. The electric field (\ref{fala18}) of the incident
wave is parallel to {\bf d}$y$. The electric field
(\ref{refl1}) of the reflected wave contains the extra component
parallel to {\bf d}$z$. Lindell and Sihvola in \cite{ismo1}
call it"cross-polarized component".

\medskip
The Poynting two-form for the reflected wave fields (\ref{refl1})
and (\ref{refl2}) reads
$${\bf S}_-(x,t)={\bf E}_-(x,t)\wedge {\bf H}_-(x,t)=-
\frac{\sqrt{\varepsilon /\mu}}{(\varepsilon +\alpha ^2\mu)^2}\,
[(\varepsilon -\alpha\mu)^2+4\alpha ^2\varepsilon \mu ]\,
E_{y+}^{~2}(\eta _+)\,{\bf d}y\wedge {\bf d}z$$
$$=-\sqrt{\varepsilon /\mu}\,E_{y+}^{~2}(\eta _+)\,{\bf d}y\wedge {\bf d}z.$$
This ought to be compared with the Poynting two-form of the
incident wave (\ref{fala18}) and (\ref{fala20})
$${\bf S}_+(x,t)={\bf E}_+(x,t)\wedge {\bf H}_+(x,t)=
\sqrt{\varepsilon /\mu}\,E_{y+}^{~2}(\eta _-)
\,{\bf d}y\wedge {\bf d}z.$$
We see that ${\bf S}_-$ is opposite to ${\bf S}_+$ on the interface
$x=0$. This implies that
the reflection coefficient $T=|{\bf S}_-|/|{\bf S}_+|$ is precisely one, what
is to be expected because no energy can be transmitted into PEMC medium.

\medskip
%It follows from (\ref{war2}) and (\ref{war3}) that
%$$f_2+g_2=-\alpha \sqrt{\frac{\mu}{\varepsilon}}\,(f_1+g_1).$$
%We separate this into two equalities
%\begin{equation} f_2=-\alpha \sqrt{\frac{\mu}{\varepsilon}}\,f_1,~~~~
%g_2=-\alpha \sqrt{\frac{\mu}{\varepsilon}}\,g_1. \label{war7}\end{equation}

We substitute now $E_{y-}$ and $E_{z-}$ known from (\ref{war6}) into
(\ref{war1}), (\ref{war2}) in order to express $f_i+g_i$
by the component of the incident wave:
\begin{equation} f_1(\omega t)+g_1(\omega t)=\frac{2\varepsilon}{\varepsilon +\alpha ^2\mu}\,
E_{y_+}(\omega t), \label{war7}\end{equation}
\begin{equation} f_2(\omega t)+g_2(\omega t)=-\frac{2\alpha \sqrt{\varepsilon \mu}}
{\varepsilon +\alpha ^2\mu }\,E_{y+}(\omega t).
\label{war8}\end{equation} We insert $g_1,\,g_2$ calculated from
these equation into (\ref{fala26}, \ref{fala28})
$$\tilde{\bf E}(x,t)=\left[ f_1(\xi_+)+
\frac{2\varepsilon}{\varepsilon +\alpha ^2\mu}\,E_{y+}(\xi
_-)-f_1(\xi _-)\right] {\bf d}y~~~~~~~~~~~~~~~~~~~~~~~~~$$
\begin{equation} ~~~~~~~~+\left[ f_2(\xi _+)-\frac{2\alpha \sqrt{\varepsilon
\mu}}{\varepsilon +\alpha ^2\mu}\,E_{y+}(\xi _-)-f_2(\xi _-)\right]
{\bf d}z, \label{konc1}\end{equation}
$$\tilde{\bf B}(x,t)=\frac{1}{\omega}\,{\bf d}x\wedge \left\{
\left[ \tilde{k}f_1(\xi _+)-\frac{2\varepsilon
\tilde{q}}{\varepsilon +\alpha ^2\mu}\,E_{y+}(\xi
_-)+\tilde{q}f_1(\xi _-)\right] \right. {\bf d}y ~~~~~~~~~~$$
\begin{equation} ~~~~~~~~~~~~~~~~~+\left. \left[ \tilde{k}f_2(\xi_+)+\frac{2
\alpha\sqrt{\varepsilon \mu}\,\tilde{q}}{\varepsilon +\alpha
^2\mu} \,E_{y+}(\xi _-)+\tilde{q}f_2(\xi _-)\right] {\bf
d}z\right\} \label{konc2}\end{equation} The other fields are
obtained through $\tilde{\bf H}=-\alpha \tilde{\bf E}$,
$\tilde{\bf D}=\alpha \tilde{\bf B}$. The arbitrary functions
$f_1,f_2$ are still present in these formulas, therefore the
electromagnetic wave in PEMC, after fulfilling the boundary
conditions, remains arbitrary to high degree.

If one chooses $f_1=f_2=0$, the formulas (\ref{konc1},
\ref{konc2}) reduce to $$ \tilde{\bf E}(x,t)=\frac{2E_{y+}(\xi
_-)}{\varepsilon +\alpha ^2\mu}\,(\varepsilon \,{\bf d}y-\alpha
\sqrt{\varepsilon \mu}\,{\bf d}z),$$ $$\tilde{\bf B}(x,t)=-
\frac{2\tilde{q}E_{y+}(\xi _-)}{\varepsilon +\alpha ^2\mu}~{\bf
d}x\wedge (\varepsilon \,{\bf d}y-\alpha \sqrt{\varepsilon
\mu}\,{\bf d}z),$$ which show that only the left-travelling wave
in present in PEMC. On the other hand, if one chooses
$$ f_1(\omega t)=\frac{2\varepsilon}{\varepsilon +\alpha
^2\mu}\,E_{y+}(\omega t),~~~~f_2(\omega t)=-\frac{2\alpha
\sqrt{\varepsilon \mu}}{\varepsilon +\alpha ^2\mu}\,E_{y+}(\omega
t),$$ the formulas (\ref{konc1}, \ref{konc2}) assume the form
$$ \tilde{\bf E}(x,t)=\frac{2E_{y+}(\xi
_+)}{\varepsilon +\alpha ^2\mu}\,(\varepsilon \,{\bf d}y-\alpha
\sqrt{\varepsilon \mu}\,{\bf d}z),$$ $$\tilde{\bf B}(x,t)=-
\frac{2\tilde{k}E_{y+}(\xi _+)}{\varepsilon +\alpha ^2\mu}~{\bf
d}x\wedge (\varepsilon {\bf d}y-\alpha \sqrt{\varepsilon
\mu}\,{\bf d}z),$$ and in this case only the right-travelling wave
is present in PEMC.

\section{Plane wave in PEMC slab}

The reasoning of previous section is based on the assumption that
PEMC medium fills the whole half-space $x>0$. This assumption is
nonphysical, it is natural to assume, rather, that the PEMC medium
forms a plane-parallel slab defined by the condition $0<x<b$. What
conditions should be imposed on the electromagnetic fields on the
other interface $x=b$? Since no energy is transmitted through
PEMC, we suppose that the same occurs for $x>b$. The zero energy
flux cannot be accomplished by a superposition of left- and
right-travelling waves giving total energy flux equal to zero,
because there is no physical reason for a presence of the
electromagnetic wave incoming from the right infinity. In this
manner we arrive to conclusion that the electromagnetic fields
should vanish for $x>b$. The boundary conditions (\ref{bound1})
are satisfied trivially because the two-forms (\ref{fala27},
\ref{fala29}) are parallel to ${\bf d}x$. The other condition
\ref{bound2}) assumes the form
$$\tilde{\bf E}(b,t)=0,~~~~~\tilde{\bf H}(b,t)=0.$$
It is sufficient to consider the first equality, because
$\tilde{H}=-\alpha \tilde{E}$, hence we assume that the two square
brackets in \ref{konc1} vanish
\begin{equation} f_1(\omega t-\tilde{k}b)-f_1(\omega t+\tilde{q}b)
+\frac{2\varepsilon}{\varepsilon+\alpha ^2\mu}\,E_{y+}(\omega
t+\tilde{q}b)=0, \label{warb1}\end{equation}
\begin{equation} f_2(\omega t-\tilde{k}b)-f_2(\omega t+\tilde{q}b)
-\frac{2\alpha \sqrt{\varepsilon \mu}}{\varepsilon +\alpha
^2\mu}\,E_{y+} (\omega t+\tilde{q}b)=0.
\label{warb2}\end{equation} Equations (\ref{warb1}, \ref{warb2})
can be rewritten with the following change of notation: $\omega
t+\tilde{q}b=u$, $\tilde{k}b+\tilde{q}b=a$, $\omega
t-\tilde{k}b=u-a$:
\begin{equation} f_1(u-a)-f_1(u)+\frac{2\varepsilon}{\varepsilon+\alpha
^2\mu}\,E_{y+}(u)=0, \label{warb5}\end{equation}
\begin{equation}  f_2(u-a)-f_2(u)-\frac{2\alpha \sqrt{\varepsilon \mu}}
{\varepsilon+\alpha ^2\mu}\,E_{y+}(u)=0
\label{warb6}\end{equation}
Thsese equations ought to be satisfied for all $u\in \real$ and fixed $a$.
They are functional equations which I do not know how to solve.

Let us assume now that the incident wave is time-harmonic, i.e.
\begin{equation} E_{y+}(u)=A\cos (u-\delta ), \label{harmo}\end{equation}
with a given constant $A$. Then we look for the unknown function $f_1$ in
the form
\begin{equation} f_1(u)=C\cos u \label{warb7}.\end{equation}
The constants $C$ and $\delta $ are to be found.

\medskip
We rewrite equation (\ref{warb5}) with the use of (\ref{harmo},
\ref{warb7}):
$$ C\cos (u-a)-C\cos u+\frac{2\varepsilon A}{\varepsilon +\alpha
^2\mu}\cos(u-\delta)=0.$$ A simple trigonometry yields
$$\left( C\cos a-C+\frac{2\varepsilon A}{\varepsilon +\alpha ^2\mu}\, \cos
\delta \right) \cos u+\left( C\sin a+\frac{2\varepsilon
A}{\varepsilon +\alpha \mu}\, \sin \delta \right) \sin u=0.$$ Sine
and cosine are linearly independent functions, thence
$$C\cos a-C+\frac{2\varepsilon A}{\varepsilon +\alpha ^2\mu}\, \cos
\delta =0.$$
$$C\sin a+\frac{2\varepsilon A}{\varepsilon +\alpha
\mu}\, \sin \delta =0.$$ This system of two equations has two
solutions:
$$C_I=\frac{\varepsilon A}{(\varepsilon +\alpha
\mu )\sin (a/2)},~~~~\delta _I=\frac{a-\pi}{2},$$
$$C_{II}=-\frac{\varepsilon A}{(\varepsilon +\alpha
\mu )\sin (a/2)},~~~~\delta _{II}=\frac{a+\pi}{2}.$$ Then,
according to (\ref{warb5}, \ref{warb6}), we have two
possibilities. The first one reads
\begin{equation}
E_{y+}(u)=A\cos \left( u-\frac{a}{2}+\frac{\pi}{2}\right) =-A\sin
\left( u-\frac{a}{2}\right) ,\label{warb8}\end{equation}
\begin{equation} f_1(u)=\frac{\varepsilon A}{(\varepsilon +\alpha
\mu )\sin (a/2)}\,\cos u. \label{warb9}\end{equation} The other
possibility is not essentially different -- it gives only opposite
sign in front of A.

A similar reasoning applied to equation (\ref{warb6}) leads to the
result
\begin{equation} f_2(u)=-\frac{\alpha \sqrt{\varepsilon \mu}\,A}{(\varepsilon +\alpha
\mu )\sin (a/2)}\,\cos u. \label{warb10}\end{equation} We
substitute (\ref{warb8}, \ref{warb9}, \ref{warb10}) into
(\ref{konc1}, \ref{konc2}) and obtain the following time-harmonic
plane electromagnetic wave in the plane-parallel slab $0<x<b$ of
PEMC medium
$$\tilde{\bf E}(x,t)=\frac{A}{\varepsilon +\alpha
^2\mu}\left[ \frac{\cos (\omega t-\tilde{k}x)}{\sin
a/2}-2\sin(\omega t+\tilde{q}x-a/2)\right. ~~~~~~~~~~~~~~~~$$
\begin{equation} ~~~~~~~~~~~~~~~~~~\left. -\frac{\cos(\omega t+\tilde{q}x)}
{\sin a/2}\right] (\varepsilon \,{\bf d}y-\alpha \sqrt{\varepsilon
\mu}~{\bf d}z), \label{konc3}\end{equation}
$$\tilde{\bf B}(x,t)=
\frac{A}{(\varepsilon +\alpha ^2\mu )\omega }\left[
\frac{\tilde{k}\cos (\omega t-\tilde{k}x)}{\sin
a/2}-2\tilde{q}\sin(\omega t+\tilde{q}x-a/2)\right. ~~~~~~~~$$
\begin{equation} ~~~~~~~~~~~~~~~~~~~~~~~~\left.-\frac{\tilde{q}\cos(\omega t
+\tilde{q}x)} {\sin a/2}\right] {\bf d}x\wedge (\varepsilon \,{\bf
d}y-\alpha \sqrt{\varepsilon \mu}~{\bf d}z),
\label{konc4}\end{equation} where $a=\tilde{k}b+\tilde{q}b$. As is
visible from (\ref{warb8}, \ref{warb9}), the values $a=2n\pi $ for
integer $n$ are not allowable for the time-harmonic wave. This
fact can be interpreted as follows. As mentioned earlier, the wave
numbers $\tilde{k},\,\tilde{q}$ are independent of the wave number
$k$ in the left ordinary medium, i.e. in the half-space $x<0$. The
values of $\tilde{k}$ and $\tilde{q}$ which yield
$\tilde{k}b+\tilde{q}b=2n\pi $ can not be present in the solution
(\ref{konc3}, \ref{konc4}) valid for the PEMC slab.

\section{Conclusion}

A plane electromagnetic wave propagating in PEMC has been
presented. The normal reflection of plane electromagnetic wave
from the PEMC boundary has been considered with the use of
boundary conditions. It turned out that the field strengths of the
reflected wave are the same as in \cite{ismo1}.

An interesting result is that the wave present in PEMC may contain
two arbitrary functions $f_1,\,f_2$ which are present in the
right- and left-travelling waves. By and appropriate choice of
them one can assure the presence of the right-travelling wave
only, or the left-travelling one only. This is possible because no
energy is transported by these waves. In a sense they are virtual
waves. Moreover, the two oppositely travelling waves in PEMC may
contain wave numbers $\tilde{k}$ and $\tilde{q}$ in their phases,
which may be different from each other and from the wave number
$k$ in the ordinary medium.

The above mentioned observations are true for the PEMC medium
extending from $x=0$ to infinity. Since this is nonphysical
situation, we have assumed that PEMC extends from $x=0$ to $x=b$,
i.e. it forms a plane-parallel slab. On a one side of it, there
are incident and reflected waves, whereas on the other side of it,
there is no electromagnetic fields at all. We have shown that the
functions $f_1,\,f_2$ must have unique shapes for the
time-harmonic wave incident from the left.

\vspace{8mm} {\bf Acknowledgement}

\medskip
A principal part of the work has been done during my stay at
Cologne University with the financial support of the European
Union. I am deeply grateful to Friedrich Hehl and Yuri Obukhov for
stimulating suggestions and discussions which make this paper
possible to emerge.

\end{document}